\documentclass[sigconf]{acmart}

\copyrightyear{2022}
\acmYear{2022}
\setcopyright{acmlicensed}
\acmConference[WiSec '22] {Proceedings of the 15th ACM Conference on Security and Privacy in Wireless and Mobile Networks}{May 16--19, 2022}{San Antonio, TX, USA.}
\acmBooktitle{Proceedings of the 15th ACM Conference on Security and Privacy in Wireless and Mobile Networks (WiSec '22), May 16--19, 2022, San Antonio, TX, USA}
\acmPrice{15.00}
\acmISBN{978-1-4503-9216-7/22/05}
\acmDOI{10.1145/3507657.3528554}

\usepackage{xcolor}
\usepackage[binary-units=true,detect-weight=true,detect-family=true]{siunitx}
\usepackage{xspace}
\usepackage{tikz}
\usepackage[inline]{enumitem}

\usepackage{soul}

\usepackage{booktabs}
\usepackage{multirow}
\usepackage{amsmath}

\usepackage[ruled]{algorithm2e}
\SetKwComment{Comment}{$\triangleright$\ }{}
\usepackage[noend]{algpseudocode}

\usepackage{units}

\definecolor{darkgrey}{RGB}{80,80,80}
\definecolor{lightgrey}{RGB}{170,170,170}

\definecolor{brown}{HTML}{a52a2a}
\definecolor{darkcyan}{HTML}{0a888a}

\newcommand{\ie}{\textit{i.e.},\xspace}
\newcommand{\eg}{\textit{e.g.},\xspace}
\newcommand{\cf}{\textit{cf.}\xspace}

\newcommand{\code}[1]{\texttt{#1}}

\newcommand{\CUT}[1]{{}}

\usepackage{acronym}

\acrodef{cps}[CPS]{Cyber-Physical System}
\acrodef{mac}[MAC]{Message Authentication Code}

\newcommand{\name}{BP-MAC\xspace}

\usepackage{subcaption}
\usepackage{tikz}

\usetikzlibrary{fit,calc}
\newcommand*{\tikzmk}[1]{\tikz[remember picture,overlay,] \node (#1) {};\ignorespaces}
\newcommand{\boxit}[1]{	
	\tikz[remember picture,overlay]{\node[yshift=3pt,fill=#1,opacity=.25,fit={(A)($(B)+(.92\linewidth,.8\baselineskip)$)}] {};}\ignorespaces
}

\definecolor{blue}{RGB}{0, 35, 161}
\definecolor{orange}{RGB}{240, 100, 0}

\definecolor{blue2}{RGB}{191, 200, 231}
\definecolor{orange2}{RGB}{251, 216, 191}

\newlength{\overwritelength}
\newlength{\minimumoverwritelength}
\newcommand{\overwrite}[4][red]{%
	\settowidth{\overwritelength}{$#3$}%
	\ifdim\overwritelength<\minimumoverwritelength%
	\setlength{\overwritelength}{\minimumoverwritelength}\fi%
	\stackrel
	{%
		\begin{minipage}{\overwritelength}%
			\color{#2!60}\centering\small #4\\%
	\end{minipage}}
	{\colorbox{#1}{\color{black}$\displaystyle#3$}}}

\newif\ifmanualheader

\manualheaderfalse

\ifmanualheader
  \makeatletter
  \if@ACM@anonymous
  \else
    \g@addto@macro\@subtitlenotes{\global\let\authors=\cleanauthors}
  \fi
  \makeatother
\fi

\begin{document}
\fancyhead{}

\title{BP-MAC: Fast Authentication for Short Messages}

\ifmanualheader
  \author{Eric Wagner$^{*,\dagger}$, Martin Serror$^{*}$, Klaus Wehrle$^{\dagger}$, and Martin Henze$^{\ddagger,*}$}
  \def\cleanauthors{Eric Wagner, Jan Bauer, Klaus Wehrle, and Martin Henze}
  \affiliation{
    $^*$\textit{Cyber Analysis \& Defense}, Fraunhofer FKIE \country{Germany} $\cdot$ \{firstname.lastname\}@fkie.fraunhofer.de\\
    $^\dagger$\textit{Communication and Distributed Systems}, RWTH Aachen University \country{Germany} $\cdot$ \{lastname\}@comsys.rwth-aachen.de\\ %
    $^\ddagger$\textit{Security and Privacy in Industrial Cooperation}, RWTH Aachen University \country{Germany} $\cdot$ henze@cs.rwth-aachen.de
  }
\else
  \author{Eric Wagner}
  \email{eric.wagner@fkie.fraunhofer.de}
  \affiliation{%
    \institution{Fraunhofer FKIE}
    \country{}
  }
  \affiliation{%
    \institution{RWTH Aachen University}
    \country{}
  }

  \author{Martin Serror}
  \email{martin.serror@fkie.fraunhofer.de}
  \affiliation{%
    \institution{Fraunhofer FKIE}
    \country{}
  }

  \author{Klaus Wehrle}
  \email{wehrle@comsys.rwth-aachen.de}
  \affiliation{%
    \institution{RWTH Aachen University}
    \country{}
  }
  \affiliation{%
	\institution{Fraunhofer FKIE}
	\country{}
}

  \author{Martin Henze}
  \email{henze@cs.rwth-aachen.de}
  \affiliation{%
    \institution{RWTH Aachen University}
    \country{}
  }
  \affiliation{%
    \institution{Fraunhofer FKIE}
    \country{}
  }
\fi

\renewcommand{\shortauthors}{\cleanauthors}

\begin{abstract}
Resource-constrained devices increasingly rely on wireless communication for the reliable and low-latency transmission of short messages.
However, especially the implementation of adequate integrity protection of time-critical messages places a significant burden on these devices.
We address this issue by proposing \name, a fast and memory-efficient approach for computing message authentication codes based on the well-established Carter-Wegman construction. 
Our key idea is to offload resource-intensive computations to idle phases and thus save valuable time in latency-critical phases, \ie when new data awaits processing.
Therefore, \name leverages a universal hash function designed for the \emph{bitwise preprocessing} of integrity protection to later only require a few XOR operations during the latency-critical phase.
Our evaluation on embedded hardware shows that \name outperforms the state-of-the-art in terms of latency and memory overhead, notably for small messages,  as required to adequately protect resource-constrained devices with stringent security and latency requirements.

\end{abstract}



\maketitle

\section{Introduction}
\label{sec:intro}

\acp{cps}, such as industrial control systems, power grids, or smart transportation, depend on mission-critical machine-to-machine communication, where short sensor and control messages need reliable transmission within a few milliseconds and below~\cite{ 2012_galloway_ics,LPDz17}. 
The wireless exchange of these messages is extremely challenging, despite being as short as a single byte~\cite{2012_galloway_ics}, as it has to account for a shared and error-prone transmission medium~\cite{2018_raza_wirelessindustrial}.
Implementing adequate security for such critical communication further impedes reaching the latency requirements due to the high processing times of cryptographic algorithms on resource-constrained devices~\cite{2018_hiller_antedated,BBGT16}.
While combining precomputed keystreams with message payload hardly introduces any communication latency for encryption~\cite{2018_hiller_antedated}, data authenticity and integrity cannot be solved as easily.
However, the lack of integrity protection in these scenarios facilitates attacks with severe consequences, ranging from financial losses to threats to human lives~\cite{BBGT16}.

A well-established method for authenticity and integrity protection is a \ac{mac}: 
By appending authentication tags (short: \textit{tags}) to each message, the sender enables the receiver to verify that a message has not been altered during transmission~\cite{2020_boneh_crypto}.
However, as these tags depend on the transmitted message, their computation and verification introduce delays during the latency-critical phase, \ie after new data becomes available at the sender and before the receiver can process it.
Since \acp{cps} rely on resource-constrained special-purpose devices, \eg sensors and actuators, they are typically not capable of computing and verifying tags without significant delay~\cite{BBGT16}.
Thus, the delay introduced by \acp{mac} must be reduced to enable secure wireless communication even under the most stringent latency requirements.

Therefore, we propose \emph{Bitwise Precomputable \ac{mac}}~(\name) to speed up integrity protection by bitwise taking advantage of the characteristics of \ac{cps} scenarios:
While messages are often only a few bytes long, special-purpose devices are usually idling until new data has to be sensed and transmitted or received and processed~\cite{2018_hiller_antedated, 2018_ankele_mergeMAC}.
Hence, \name offloads message-independent and resource-intensive computations to a preprocessing phase where devices are otherwise idle.
Moreover, as a Carter-Wegman construction~\cite{1979_carter_universal,1981_wegman_new}, \name's security can be reduced to traditional time-proven cryptographic primitives and can be trusted to protect even the most sensitive and critical communication against manipulations.

\textbf{Contributions.} We thus present the following contributions:
\begin{itemize}[noitemsep,topsep=0pt,leftmargin=9pt]
	\item We propose \name, a novel Carter-Wegman scheme requiring only a few XOR operations during the latency-critical phase. 
	It precomputes and caches authentication data for each bit of a message and securely combines them to compute new tags.
	\item Since caching comes with a significant memory trade-off, we introduce memory optimizations realizing a smaller memory footprint than \name's closest contender (UMAC~\cite{2006_krovetz_umac}) for messages smaller than 12 to 43 bytes, depending on tag lengths.
	\item We evaluate \name's performance on different processor architectures using the Zolertia RE-Mote and Z1 boards to show \name's capability to reduce the overhead of integrity protection by more than an order of magnitude for small messages.
\end{itemize}
\vspace*{1mm}

\textbf{Availability Statement.}
The source code underlying this paper
is available at: \url{https://github.com/fkie-cad/bpmac}

\section{Message Authentication Codes}
\label{sec:macs}
In the following, we formally introduce \acfp{mac} (Sec.~\ref{sec:macs:formal}) and universal hashing~(Sec.~\ref{sec:macs:carter-wegman}) as a basis for efficient, secure authentication schemes.
Then, we discuss related work on optimizing \acp{mac} for low-latency scenarios (Sec.~\ref{sec:macs:rw}).

\subsection{Formal Definition}
\label{sec:macs:formal}

A \ac{mac} scheme is composed of two algorithms, \textit{Sig}$_k$ and \textit{Vrfy}$_k$, that generate and verify authentication tags $t$ with the help of a pre-shared secret key $k$~\cite{2020_boneh_crypto}.
A message $m$, extended by a tag $t$, computed as $t=$ \textit{Sig}$_k(m)$, allows the receiver of a message to verify the integrity of the received message.
Upon reception of a message-tag pair $(m,t)$, the receiver uses the algorithm \textit{Vrfy}$_k(m,t)$ to verify whether the message or the tag has been manipulated in transit by an attacker not possessing the secret key $k$.
In most popular \ac{mac} schemes, \textit{Vrfy}$_k(m,t)$ computes the tag $t^*=$ \textit{Sig}$_k(m)$ based on the received message $m$ and returns whether $t^*$is identical to the received tag $t$.
Consequently, \ac{mac} schemes are considered secure if, even under a chosen-message attack, an attacker cannot create an existential \ac{mac} forgery, \ie a tag $t$ for a previously unseen message $m$, that \textit{Vrfy}$_k(m,t)$ would accept.
Achieving this property typically requires the execution of (for resource-constrained devices) processing-intensive cryptographic algorithms.

\subsection{The Carter-Wegman MAC Construction}
\label{sec:macs:carter-wegman}

The Carter-Wegman construction allows building efficient and secure \ac{mac} schemes, such as UMAC~\cite{2006_krovetz_umac} and Poly1305~\cite{2005_bernstein_poly1305}, which gained popularity due to their superior performance compared to traditional schemes.
Carter-Wegman \ac{mac} constructions add an additional nonce $n$ to the parameters of \textit{Sig}$_k$ and \textit{Vrfy}$_k$, which natively protects against replay attacks~\cite{2020_boneh_crypto}.
These constructions first compute the digest of the message $m$ with a hash function unknown to the attacker.
Then, this digest is masked by the encrypted nonce.
More formally, a Carter-Wegman \ac{mac} construction computes the tag $t_i$ for a message $m_i$ and a nonce $n_i$ as $t_i = F_{k1}(m_i) \oplus H_{k2}(n_i)$, where $H$ is a pseudorandom function covering the same output space as the universal hash function $F$~\cite{2020_boneh_crypto}.
A universal hash function is a (not necessarily cryptographic) secret hash function for which it is hard to find a collision, \ie finding $m, m'$ such that $F(m)=F(m')$, as long as the attacker does not know a single output of $F$~\cite{2020_boneh_crypto}.

\subsection{Efficient MACs for Low Latency Scenarios} %
\label{sec:macs:rw}

Optimized computations of established primitives such as AES or SHA256 have been proposed to address the need for fast message authentication on resource-constrained devices.
Such approaches include hardware acceleration~\cite{2017_yang_hardware-survey} or the preprocessing of authentication for predictable parts of future messages~\cite{2018_hiller_antedated}.
However, these approaches still fail to enable secure sub-millisecond communication as many \ac{cps} scenarios demand~\cite{ 2012_galloway_ics,LPDz17}. 
Consequently, a range of novel \ac{mac} algorithms based on new cryptographic primitives has been proposed to reduce processing and latency overheads for resource-constrained devices, including SipHash~\cite{2012_aumasson_siphash}, TuLP~\cite{2014_gong_tulp}, Chaskey~\cite{2014_mouha_chaskey}, and LightMAC~\cite{2016_luykx_lightmac}.
MergeMAC~\cite{2018_ankele_mergeMAC} saves valuable time during latency-critical phases by extracting predictable parts of future messages to authenticate them in advance.
However, mission-critical \ac{cps} must rely on scrutinized and time-proven cryptographic primitives to ensure the maximum possible security guarantees. 

Therefore, competing approaches target efficient Carter-Wegman MAC schemes, whose security can be reduced to underlying cryptographic primitives~(\eg~AES).
To do so, UMAC~\cite{2006_krovetz_umac} and VMAC~\cite{2006_krovetz_vmac} use custom hash functions optimized for 32-bit and 64-bit architectures, respectively.
The universal hash function employed by Poly1305 evaluates a polynomial with the message as coefficients over the finite field $\mathbb{Z}_{2^{130} -5}$~\cite{2005_bernstein_poly1305}.
Although highly efficient, these schemes optimize for messages of at least a few hundred bytes~\cite{2012_aumasson_siphash}.
PMAC~\cite{2002_black_pmac} fragments messages to parallelize and thus speed up tag computations. 
Nevertheless, typical resource-constrained devices with single-core processors do not benefit from such optimizations.

Another branch of research tackles the reduction of bandwidth overhead introduced by \acp{mac} schemes via progressive \acp{mac}~(ProMACs~\cite{2020_armknecht_promac}).
ProMACs reduce the tag sizes by aggregating authentication over several messages~\cite{2020_armknecht_promac, 2020_li_cumac, 2021_li_cumacs, 2017_schmandt_minimac, spmac}.
Thereby, ProMACs protect each message with immediately reduced security, allowing their optimistic processing without waiting for the reception of subsequent messages.
Still, a message's authenticity is reinforced with subsequent messages to ensure eventual strong security~\cite{2020_armknecht_promac}.
Like traditional authentication, ProMACs also benefit from faster underlying \ac{mac} schemes to reduce processing latency.

\section{BP-MAC: A Bitwise Precomputed MAC}

To remove the bottleneck of cryptographic processing for message authentication and thus enable low-latency transmissions of small messages on resource-constrained devices using conventional security primitives, we propose \emph{Bitwise Precomputable \ac{mac}} (\name).
\name combines the idea of preprocessing message-independent computations with a Carter-Wegman \ac{mac} construction specifically designed for small messages.
We begin by discussing preprocessing as the underlying principle of \name~(Sec.~\ref{sec:design:idea}) before detailing \name's construction~(Sec.~\ref{sec:design:details}).
Then, we present memory optimizations~(Sec.~\ref{sec:design:implementation}) and discuss the security of \name{}~(Sec.~\ref{sec:design:security}).

\subsection{ Precomputations for Fast Authentication}
\label{sec:design:idea}

\name takes advantage of precomputations since the processing loads in \ac{cps} scenarios are not evenly distributed over time~\cite{2018_hiller_antedated, 2018_ankele_mergeMAC}.
Instead, devices periodically have idle times until new data has to be transmitted, received, or processed.
Hence, the idea behind \name is to offload computationally expensive operations into those idling phases to reduce computations during latency-critical phases to a minimum.
Therefore, a naïve approach would be to precompute and store authentication tags for each possible message.
However, precomputing authentication information, \eg for all possible, rather tiny two-byte messages would then already consume 1\,MB of memory~(assuming the recommended tag length of 16\,bytes), a number that is exponentially growing for longer messages.
Hence, this naïve approach is infeasible for resource-constrained CPS devices.

In contrast, the core idea of \name is to precompute tags for individual bits, which are then efficiently combined when a message needs authentication.
As each bit can only represent either 0 or 1, \name's memory consumption grows linearly with message lengths and amounts to only 512\,bytes for \emph{all} possible two-byte long messages.
These tags can be precomputed once a new key is exchanged or even provided by the (more powerful) communication partner.
In addition, \name masks the combined tags with a secret masking tag to prevent replay attacks and enable their secure aggregation.
Such tags can be conveniently precomputed between transmissions since they are independent of the actual message.

\subsection{ \name's Design in Detail}
\label{sec:design:details}

\begin{figure}
	\centering	
	\includegraphics[width=\columnwidth]{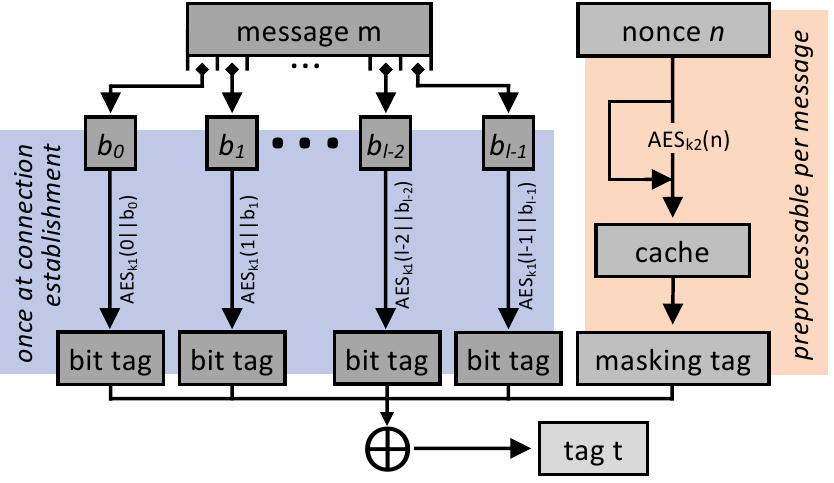}
	\vspace*{-6mm}
	\caption{
		\name achieves high-speed tag computations during the latency-critical phase by only XORing the bit tag depending on each bit value and the masking tag to generate a secure authentication tag.
		The masking tag is independent of the message and can thus be precomputed during idle times.
		Bit tags are precomputed once for each unique key.
	}
	\vspace*{-3mm}
	\label{fig:construction}
	
\end{figure}

\name achieves fast authentication by only executing the efficient aggregation of previously precomputed tags during the latency-critical phase.
We illustrate the different steps of \name's design in Figure~\ref{fig:construction} and discuss them in detail in the following.

\subsubsection{Bit Tags}

As shown in Figure~\ref{fig:construction}, bit tags are intermediate tags computed for each message bit.
To ensure \name's security and prevent collisions, all bit tags have to be a unique pseudorandom number that must be kept secret.
\name computes bit tags as the digest of a pseudorandom function by concatenating bit index and bit value.
Concretely, \name uses AES-128$_{k1}$ as a pseudorandom function.
Here, $k1$ is a secret key shared between both communicating parties due to its increased efficiency over other functions, \eg HMAC-SHA256.
Thus, if the value of the third bit in the second byte is zero, the corresponding bit tag is AES-128$_{k1}(10||0)$.
Overall, there exist $2\cdot{}l$-bit tags that have to be precomputed for a message with $l$ bits.
However, this only occurs once a new key is established and could even be offloaded to a more powerful device if necessary.

\subsubsection{Masking Tags}

The purpose of masking tags is twofold.
First, they provide replay protection by incorporating a nonce into the final tag.
Second, these pseudorandom tags prevent the leakage of information concerning individual bit tags and are thus crucial for the security of \name.
As nonce, \name uses a zero-initialized counter that is incremented for each newly computed tag.

Computing the masking tag from a pseudorandom function and the nonce is analogous to UMAC~\cite{2006_krovetz_umac} and relies on AES-128$_{k2}$.
For tags with a length between 9 and 16 bytes, we use the leading $n$ bytes of the AES-encrypted nonce, where $k2$  $(\neq k1)$ is also a secret key shared between both parties.
For smaller tags, we use the cached 16-byte long AES output for multiple masking tags, as long as no output bytes are reused.
Thus, for tags of 4 bytes, only every fourth nonce has to be encrypted to reduce processing overhead, while we can use the unused bytes of the cached output to blind the remaining tags.
As nonces can be predicted, masking tags can be precomputed and thus do not add to the latency-critical delay.

\subsubsection{Tag Computation and Verification}

After discussing the two precomputation phases generating the bit and masking tags, we now explain the computation of the $n$-th authentication tag $t_n$ for the message $m_n$.
As shown in Figure~\ref{fig:construction}, the final step of tag computation consists of XORing the intermediate tags.
This observation also manifests in the formal definition of computing tag $t_n$:
\begin{equation*}
       \centering
       \textrm{t\textsubscript{n}} =  \bigoplus_{i \in |m_n|}  \overwrite[blue2]{blue}{\textrm{AES-128\textsubscript{k1}}(i || m_n[i])}{bit tags}\,\oplus\,\overwrite[orange2]{orange}{\textrm{AES-128\textsubscript{k2}}(n)}{masking tag}
\end{equation*}

The second part is the masking tag that is XORed with the bit tags that correspond to the message. 
Here $|m|$ denotes the length of the message and $m[i]$ denotes the value of the $i$-th bit.
Thus, all computations besides these final XOR operations are independent of the message that should be authenticated.
Therefore, \name's construction enables the quick computation of new tags, especially for short messages which require a low number of XOR operations.

As usual for deterministic MAC algorithms, the tag verification happens by computing the tag $t^*$ for the received message $m$ and then comparing it to the received tag $t$.
If both tags are identical, the receiver can conclude, with high confidence, that the received message $m$ has not been altered in transit.

\subsection{Memory Optimizations for \name}
\label{sec:design:implementation}

The naïve realization of \name requires the storage of $2\cdot|m|$ bit tags.
For a fixed-length messages, we can, however, half this memory overhead by precomputing the default tag of a message composed of only zeros.
To support the same optimization for variable-length messages, we must pad the message before authenticating it.
Then, we only need to store how to alter the tag for a bit set to one with so-called \textit{bitflip} tags, of which we need $|m|$ in total (one for each bit).
We sketch the tag computation with this optimization in Algorithm~\ref{alg:sig}, highlighting in blue the computations executed during the latency-critical phase.
Already before a new message is ready, we can compute the masking tag and XOR it with the default tag t\textsubscript{default} to generate the tag of an all-zero message as:
\begin{equation*}
	\centering
		\textrm{t\textsubscript{default}}  = \bigoplus_{ 0 \leq i < n} \textrm{AES-128\textsubscript{k1}}(i || 0)
\end{equation*}
Then, to compute the tag for the actual message, we need to change the tag for each bit that is not zero as assumed for the default tag t\textsubscript{default}.
To efficiently realize these changes, we can precompute bitflips tags for all $i$ bits in a message as:
\begin{equation*}
	\centering
    \textrm{t\textsubscript{bitflip}}[i]  =  \textrm{AES-128\textsubscript{k1}}(i || 0) \oplus \textrm{AES-128\textsubscript{k1}}(i || 1)
\end{equation*}

By XORing these bitflip tags to the default tag t\textsubscript{default}, we compute the tag as if the bit at the corresponding position is set.
Finally, we pad the message to a fixed length to be able to differentiate between a shorter message and one with trailing zeros.
\name uses the \textit{Padding method 2} as described by ISO/IEC 9797-1~\cite{ISO/IEC9797-1}, appending a 1-bit at the end of the message followed by as many 0-bits as necessary to reach the desired message length.
As the default tag t\textsubscript{default} already assumes that all bits are zero, we only need to incorporate a single further bitflip tag t for the first index after the end of the message.
Thus, we only need to store $|m|+3$ tags (one for each bit, including padding, t\textsubscript{nonce}, and t\textsubscript{default}).
This procedure reduces the number of operations during the latency-critical phase for each 0-bit, which accounts for about half of encrypted traffic.

\begin{algorithm}[t]
	
	\caption{Memory-efficient signature generation \textit{Sig*\textsubscript{k}}}\label{alg:sig}

	\textbf{Input:} message $msg$, nonce $nonce$\\
	\textbf{Output:} tag t
	
	\vspace*{1mm}
	\tikzmk{A}
	t $\leftarrow$ AES\textsubscript{k2}($nonce$) \Comment{Preprocessable masking tag}
	\vspace*{1mm}
	
	\tikzmk{B}
	\boxit{orange}
	len $\leftarrow$ length(msg) \Comment{The length of the message in bits}
	\vspace*{1mm}
	
	t $\leftarrow$ t $\oplus$ t\textsubscript{default}  \Comment{ Compute tag of all-zero message }\\
	\vspace*{2mm}
	\tikzmk{A}
	\For{n between 0 and len-1}{
		\If {the n-th bit in msg is set} {
			t $\leftarrow$ t $\oplus$ t\textsubscript{bitflips}[n]  \Comment{ Lookup the cached change to t }
		}
	}
	\vspace*{1mm}
	
	t $\leftarrow$ t $\oplus$ t\textsubscript{bitflips}[len]  \Comment{ Padding for the message }
	\vspace*{1mm}

	\tikzmk{B}
	\boxit{blue}
	\textbf{return} t
\end{algorithm}

\subsection{Security Discussion}
\label{sec:design:security}

\name utilizes the secure Carter-Wegman MAC construction~\cite{1979_carter_universal,1981_wegman_new} in which a tag $t$ for a message $m$ and a nonce $n$ is computed as
$t = F_{k1}(m) \oplus H_{k2}(n)$, where $H$ is a pseudorandom function covering the same output space as the universal hash function $F$~(\cf~Sec.~\ref{sec:macs:carter-wegman}).
For $H$, \name uses the same AES-based procedure as UMAC~\cite{2006_krovetz_umac}, with the sole difference that it computes the result in advance.
Meanwhile, the fragmentation of messages into individually authenticatable bits is a special case of the established universal hash function $F^\oplus$~\cite{2020_boneh_crypto}.
Thus, \name is secure as long as the underlying cryptographic primitive~(\ie~AES) is secure.
Moreover, \name can easily exchange this primitive, if ever considered not secure enough.
In the following, we define the considered threat model and subsequently discuss possible attack scenarios.

\subsubsection{Threat Model}

The attacker's goal is to alter traffic such that the recipient of a message accepts the modified message is genuine.
To achieve this goal, the attacker can observe and alter the (plaintext) messages and exchanged tags to either learn the secret key used by BP-MAC or directly alter a message and the corresponding tag.
The information to realize such an attack can be extracted directly from the observed traffic or through side-channel information by observing the timing of individual packets.
However, we explicitly do not consider an attacker with (partial) control over one or both communicating entities that could try to access keying information through a \eg cache side-channel attacks.
However, BP-MAC does not prevent the use of common mitigation techniques that protect against such attacks~\cite{2018_lyu_survey}.

\subsubsection{Resilience to Key Recovery Attacks}

\name is not susceptible to key recovery attacks, as otherwise a key recovery attack against the underlying cryptographic primitive, \ie AES, would exist.
By overhearing transmissions, an attacker learns $t = F^{\oplus}_{k1}(m) \oplus H_{k2}(n)$ for a known message $m$ and nonce $n$.
An attacker cannot learn $k2$ from such tags, as otherwise there would exist a key recovery attack against $H_{k2}$, \ie \code{AES-128}: After learning a digest $d=H_{k2}(m)$, the attacker could generate a key $k'$, consider all unique messages $m$ as nonces, and compute a \name tag $t'$ for arbitrary new messages $m'$ as $t'=F^{\oplus}_{k'}(m') \oplus H_{k2}(m)$.
Thus, if an attack that recovers $k2$ from $t$ exists, this attack could be used to recover $k2$ from simple $H_{k2}$ digests.
Similarly, $k1$ cannot be recovered by an attacker.
Even if an attacker would learn the output of $F^{\oplus}_{k1}(m)$, a recovery of $k1$ would imply the existence of a key recovery attack against \code{AES-128}: The attacker can compute $F^{\oplus}_{k1}(m_1||\cdots||m_k)$ for observed ciphertexts $ct=AES_k(pt)$ by considering the first part of an observed plaintext $pt$ to be the index and the second part to be a truncated message, \ie $pt=i||m_i$.
Then, if a key recovery attack would exist against $F^{\oplus}$, this same attack would also enable recovering $k$ for \code{AES} plaintext-ciphertext pairs.

\subsubsection{Unforgeability of Authentication Tags}

However, a successful attack does not need to recover the authentication keys to attack a MAC scheme.
In many cases, it suffices if an attacker can forge authentication tags and thus inject harmful messages into a protected communication.
Nevertheless, \name is also secure against such attacks.
First, a message is hashed by $F^{\oplus}$, with an unpredictable result for the attacker if they did not observe previous digests of $F^{\oplus}$.
Therefore, each output of $F^{\oplus}$ is masked by a unique pseudorandom number, ensuring that an attacker cannot learn a single output of $F^{\oplus}$.
As the nonce changes for each message, no information about the internal state of \name is revealed to an outsider.
Thus, \name is also secure in the presence of an attacker that merely wants to generate valid tags for altered or new messages.

\subsubsection{Timing Side-Channel Attack against \name}

The presented memory optimizations for \name execute some computations only for bits set to one, potentially enabling timing side-channel attacks.
In particular, the time a message authentication computation takes might hint at the number of bits set to one in the authenticated message.
If the message is sent in plaintext, this information is already accessible to third parties and does not leak confidential information or help in recovering keys.
However, when authenticating encrypted traffic, it is crucial operating in the \emph{encrypt-then-MAC} mode~\cite{2020_boneh_crypto} not to reveal information about the plaintext.

\section{Performance Evaluation}
\label{sec:eval}
\name reduces the delay of message authentication for short, mission-critical wireless communication.
We validate this claim by performing measurements on two different embedded devices (Sec.~\ref{sec:eval:latency}) and evaluating \name{}'s memory consumption (Sec.~\ref{sec:eval:memory}).

\subsection{Latency Measurements}
\label{sec:eval:latency}
We implement \name for Contiki-NG and evaluate its performance on two different architectures to ensure the observed benefits generalize across them:
Zolertia Z1~(MSP430\,@\,16\,MHz, 16-bit\,CPU, 8\,kB RAM) and Zolertia RE-Mote (ARM Cortex-M3\,@\,32\,MHz, 32-bit\,CPU, 16\,kB RAM).
To assess the performance of \name, we have to compare it to other MAC schemes with similar security guarantees, such as UMAC, VMAC, or Poly1305.
We choose to compare \name against an optimized implementation of UMAC\footnote{https://fastcrypto.com/umac} as it introduces the least processing overhead for short messages (<32\,bytes) even on an unfavorable architecture~\cite{2006_krovetz_vmac}.

\subsubsection{\name on the Zolertia Z1}

\begin{figure}
	\centering 
	\includegraphics[width=\columnwidth]{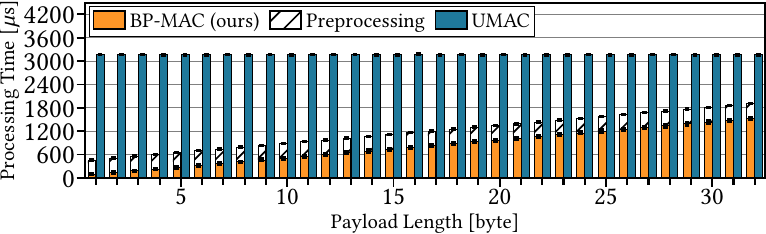}
	
		\vspace*{-2mm}
	\caption{On the 16-bit architecture of the Zolertia Z1, \name outperforms UMAC by up to two orders of magnitude for short messages and shows significantly less overall processing overhead for 32\,byte long messages.\vspace*{-3mm}}
	
	\label{fig:latency:z116}
\end{figure}

First, we compare \name's and UMAC's performance on the 16-bit MSP430 processor of the Zolertia Z1, a typical architecture for resource-constrained \ac{cps} devices. 
Here, we report on the time for computing 16-byte tags for messages of varying lengths.
Figure~\ref{fig:latency:z116} depicts the means and 99\% confidence intervals for the respective computations of one tag.
We measured 100 tag computations and derive from this the time of one tag computation due to a too low clock resolution.
We repeated each measurement 30 times.
Note that Sig$_k$ and Vrfy$_k$ require one tag computation each, such that the actual delay introduced into one secure transmission is twice the reported time.

As expected, the time required to compute UMAC tags is independent of the message length on the analyzed scale.
In contrast, \name's bitwise processing introduces a linear dependency between message length and processing time.
We observe that \name significantly outperforms UMAC, such that even for 32 bytes long messages, the overall processing time of \name is still 40\,\% lower.
For 1 byte long messages, the difference is even more extreme, as \name induces a delay of only 86 $\mu s$ for one tag computation, compared to 3.2\,ms for UMAC.
\name is thus significantly faster than UMAC for short messages on a 16-bit architecture.

We repeated the same measurements for shorter tags (12, 8, and 4\,bytes).
There, we observed similar behavior, whereas the absolute processing time of \name became shorter by $24.8-35.4\%$, $38.6-54.0\%$, and $51.1-66.5\%$  for increasingly shorter tags as shorter tags require less XOR operations. 
In turn, UMAC also becomes faster, primarily through its ability to reuse AES computations to mask multiple tags for 8 and 4-byte tags.
However, the irregular need to compute AES blocks introduces jitter to the processing of UMAC, which is undesirable in low-latency communication~\cite{2014_frotzscher_requirements}.
For \name, these bursty computations only occur in the preprocessing phase and thus do not influence actual transmission delays.

\subsubsection{\name on the Zolertia RE-Mote}

\begin{figure}
	\centering
	\includegraphics[width=\columnwidth]{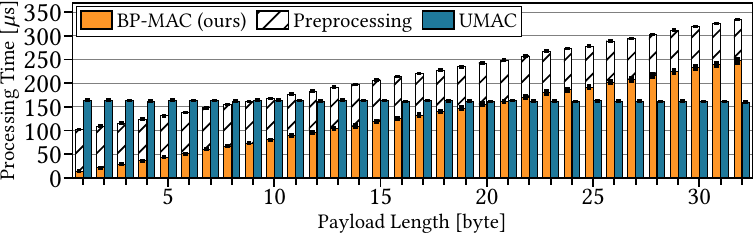}
	
	\vspace*{-2mm}
	\caption{ Even on the faster Zolertia RE-Mote, \name realizes faster tag computations than UMAC in the latency-critical phase for messages shorter than 21\,bytes and less overall processing for messages shorter than 10\,bytes.\vspace*{-3mm}} 
	
	\label{fig:latency:remote16}
\end{figure}

To show the performance of \name on a more powerful device, we furthermore compare \name and UMAC on the Zolertia RE-Mote.
While it still constitutes an embedded device, its ARM Cortex-M3 is significantly more powerful than the Zolertia Z1.
Also, its 32-bit processor is precisely the architecture targeted by UMAC.
Hence, we repeat our previous measurements and report on the results in Figure~\ref{fig:latency:remote16}.

Still, \name outperforms UMAC for small messages, partially by more than one order of magnitude.
To be precise, the latency-critical processing of \name is faster for messages not longer than 21\,bytes. 
Furthermore, even the overall processing overhead is smaller for messages shorter than 10\,bytes.
These results thus show that \name is particularly suited for authenticating short messages, even in a worst-case comparison to UMAC.
For shorter tags, this trend continues, with \name outperforming UMAC by 
$17.2-24.4\%$, $22.8-32.0\%$, and $35.2-48.3\%$
during the latency-critical phase for tags of sizes 12, 8, and 4, respectively.
The tipoff points below which \name outperforms UMAC are 26, 18, and 15-byte messages for increasingly shorter tags.

Concluding, \name enables secure communication based on the established security of AES with low processing overhead for small messages across different processor architectures.
However, the performance gap between \name and UMAC is much narrower on the Zolertia RE-Mote.
This behavior is not due to a worse performance by \name, but instead due to UMAC being specifically optimized for the 32-bit processor in this scenario.

\balance

\subsubsection{\name and Hardware-accelerated Cryptography}

Due to the rising importance of low-latency security in \ac{cps} scenarios, low-powered devices increasingly provide hardware accelerators for cryptographic operations.
However, using the Zolertia RE-Mote's sha-256 accelerator for HMAC computations shows that even hard\-ware-accelerated cryptography can introduce significant processing overheads.
Indeed, our measurements reveal that the computation of one HMAC-SHA256 tag takes approximately 220\,$\mu s$, independent of tag and payload lengths for small payloads.
Thus, in these cases, hardware-accelerated HMAC-SHA256 is even slower than a software implementation of UMAC.
Still, accelerators for specialized MAC schemes can be constructed to achieve even faster cryptographic operations.
Fittingly, \name's heavy reliance on XOR operations and its high parallelizability by individually processing each bit promise highly efficient hardware implementations.

\begin{figure}
	\centering
	\includegraphics[width=\columnwidth]{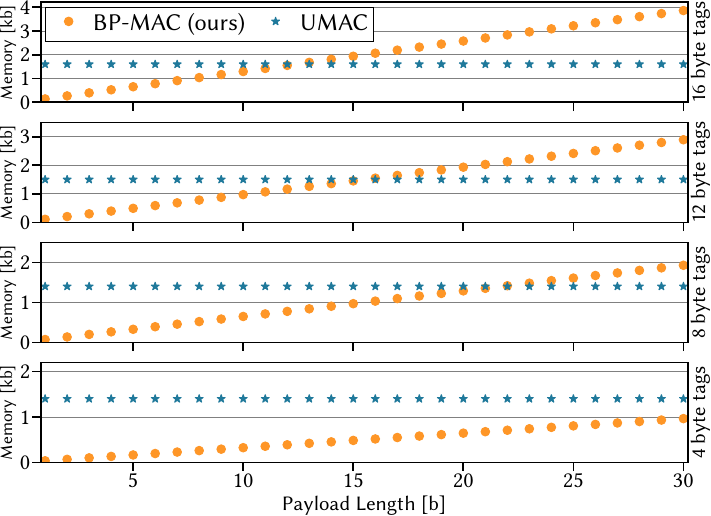}
	\caption{\name's memory footprint grows linearly with tag and message lengths. Thus, for messages shorter than 12, 15, 21, and 43 bytes, \name requires less memory than UMAC for 16, 12, 8, and 4-byte long tags, respectively.} 
	\label{fig:memory}
\end{figure}

\subsection{Memory Overhead}
\label{sec:eval:memory}
Fast MAC schemes based on universal hashing, such as \name and UMAC, typically trade memory usage for high processing speeds, a limited resource on low-power devices.
Hence, we compare \name's and UMAC's memory footprint by compiling Con\-ti\-ki-NG as a native Linux application.
We then use Valgrind's massif tool to analyze the peak memory footprint of both schemes for varying message and tag lengths. 
We depict our results in Figure~\ref{fig:memory}.

We notice that the memory footprint of UMAC is constant for different message sizes and hardly changes across tag lengths~(ranging from 1.4 kB to 1.6 kB).
In contrast, \name's memory footprint increases with the sizes of tags and messages.
Overall, \name's memory footprint increases by eight times the tag lengths when messages become one byte longer, which we expected since an additional bitflip tag has to be stored for each additional message bit.
Consequently, \name's memory footprint is favorable for small messages up until a tipoff point where UMAC becomes, in turn, more resource-efficient.
These tipoff points lie at a message size of 12, 15, 21, and 43\,bytes for 16, 12, 8, 4\,byte long tags, respectively.

Overall, our evaluation thus shows that \name outperforms UMAC in terms of processing latency and memory footprint for small messages while providing the same security guarantees, \ie reducible to the security of AES.
Thus, \name enables secure communication in critical CPS scenarios with a significantly smaller impact on communication latency than state-of-the-art approaches.

\section{Conclusion}

\acp{cps} rely on low-latency wireless communications with stringent security guarantees.
In this context, a significant challenge is to ensure the authenticity and integrity of exchanged messages, which typically leads to computationally intensive calculations in latency-critical phases, e.g., once a new message is ready for transmission.
While numerous MAC schemes aim to reduce latency, none address small messages of only a few bytes, an essential characteristic of many latency-critical CPS scenarios.
To fill this gap, we propose a new Carter-Wegman \ac{mac} scheme with AES-based security (\name), optimizing the extensive use of preprocessing with a universal hash function for the aforementioned small messages.
Thus, \name reduces latency-critical computations to a few XOR operations, which overall leads to more than a ten-fold reduction in processing overhead compared to the state-of-the-art.
Consequently, BP-MAC enables integrity protection of small messages in latency-critical scenarios, even for \ac{cps} devices with limited computational and memory resources.

\begin{acks}
Funded by the Deutsche Forschungsgemeinschaft (DFG, German Research Foundation) under Germany's Excellence Strategy -- EXC-2023 Internet of Production -- 390621612.
We thank the reviewers and our shepherd Kasper Rasmussen for their fruitful comments.
\end{acks}

\bibliographystyle{ACM-Reference-Format}
\bibliography{paper}

\end{document}